\documentclass[final,5p,times,twocolumn]{elsarticle}

\usepackage{amsmath, amssymb, amsfonts, latexsym}
\usepackage{epsfig}
\usepackage{adjustbox}
\usepackage{mathtools}
\usepackage{amssymb}
\usepackage{xcolor}
\usepackage{textcomp}
\usepackage{graphicx}
\usepackage{balance}
\usepackage{hyperref}

\journal{Nuclear Instruments and Methods}

\begin{document}
\begin{frontmatter}


\title{Precise Beam Energy Determination for Hall A after the CEBAF 12~GeV Upgrade}

\author[UNH,MIT]{S.~N.~Santiesteban}
\author[ODU,UVA]{L.~Tracy}
\author[JLab]{D.~Flay}
\author[JLab]{D.~W.~Higinbotham}
\author[in2p3]{D. Marchand}
\author[in2p3]{P. Vernin} 
\author[JLab,myfootnote]{A~Saha}

\address[UNH]{University of New Hampshire, Durham, NH 03824 USA}
\address[MIT]{Massachusetts Institute of Technology, Cambridge, MA 02139}
\address[ODU]{Old Dominion University, Norfolk, VA 23529 USA}
\address[UVA]{University of Virginia, Charlottesville, VA 22901, USA}
\address[JLab]{Jefferson Lab, Newport News, VA 23601 USA}
\address[in2p3]{Universit\'{e} Paris-Saclay, CNRS/IN2P3, IJCLab, 91405 Orsay, France}

\fntext[myfootnote]{Deceased}

\begin{abstract}
Precise and accurate measurements of the beam energy delivered to the 
experimental halls at the Thomas Jefferson Accelerator Facility (Jefferson Lab) is required by 
many experiments for proper data analysis and physics event reconstruction.  
During the 6~GeV era of Jefferson Lab, the energy delivered to experimental Hall A 
was determined to 2E-4 dE/E with multiple measurements; but after the machine 
was upgraded to 12~GeV, the accelerator's beam energy calculations needed to be re-calibrated.
In order to link the 6~GeV era calibrations to the 12~GeV era, 
the Hall A ARC energy measurement system was left unmodified.    After the upgrade, this system 
was used to determine the absolute beam energy being delivered into Hall A and find the new calibrations for the main machine.
To ensure the validity of these results, they have been cross checked using 
elastic scattering data as well as spin precession data.  
\end{abstract}


\end{frontmatter}

\section{Introduction}

The Continuous Electron Beam Accelerator (CEBA) located at the Thomas Jefferson National Accelerator Facility (Jefferson Lab) started operations in the mid-1990s to  provide a multi-GeV continuous-wave (CW) electrons beam to three experimental halls~\cite{Leemann:2001dg,Grames:2010zz}. 
For the experimental nuclear physic program, it was crucial to determine the absolute energy of the CEBAF beam both precisely and accurately for the absolute cross section measurements.   Two dedicated systems were used to determine the energy of the beam being delivered into experimental Hall A: 
\begin{itemize}
    \begin{item}
    ARC Energy: Mapping the field of the eight dipoles which bend the CEBAF beam into experimental hall A by using using a 9th dipole that is powered in series with the other eight but is outside the radiation area and can be continuously monitored~\cite{828507,marchand:tel-00298382}.
    \end{item}
    \begin{item}
    eP:   Using elastic scattering measurements with a dedicated system to precisely measure scattering angles to absolutely determine the beam energy~\cite{Berthot:1999jp}. 
    \end{item}
\end{itemize}
Another way to determine the energy involved making measurements 
of the degree of longitudinal electron polarization at the source and 
in the experimental halls~\cite{Grames:2004mk,Higinbotham:2009ze,Higinbotham:2013pgc}.



%
%

The completion of the 12 GeV upgrade at Jefferson Lab arrived with a full 
set of new experiments that must have a precise knowledge of the beam energy
for the physics analysis.  In general, Hall A requires a determination 
of the absolute beam energy of dE/E = 10$^{-4}$. The current energy
measurements is done using the Arc method \cite{Berthot},  named after the
location where the hardware is placed. It uses the section of the beamline
that connects the accelerator and the Hall (the ARC). This method was first
developed by the French collaboration of Pascal Vernin and the Scalay group
\cite{Berthot}, and it uses the ARC as a spectrometer to calculate the exact
value of the energy. Table ~\ref{Tab:beamparameters} summarizes the beam
energy parameters delivered to Hall A.

\begin{figure}[ht]
  \includegraphics[width=0.45\textwidth]{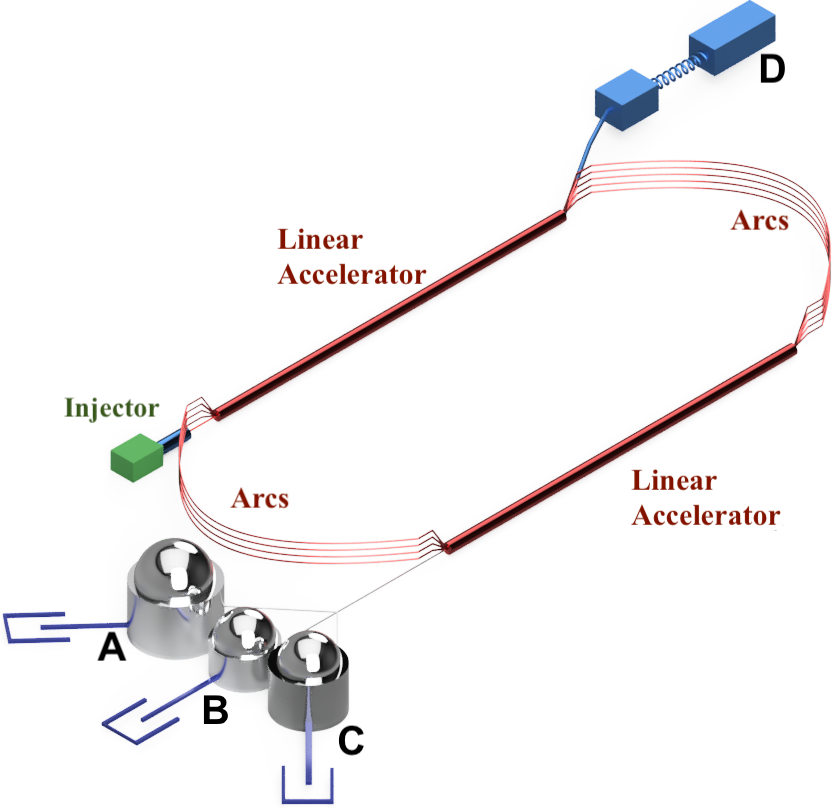}
  \caption{CEBAF schematic, the injector sends the beam to two linear accelerators (linacs), which deliver the beam to the Halls A, B, C or D.}
  \label{fig:cebaf}
\end{figure}

\begin{table}[ht]
\begin{center}
\vspace{0.5cm}
  \begin{tabular}{c c}
  \hline
 \textbf{Beam Property} &  \textbf{Nominal Value (Range)} \\ \hline 
Current       & 1-120 $\mu$A         \\ \hline                
Beam polarization   & up to 85 $\%$  \\ \hline               
Energy Spread & 
\begin{tabular}{c c c c c}
First  && $<$ 10$^{-4}$       \\
Second && $<$ 10$^{-4}$       \\
Third  && $<$ 10$^{-4}$       \\
Fourth && 3$\times$10 $^{-4}$ \\
Fifth  && 5$\times$10 $^{-4}$ \\
\end{tabular}\\  
\hline
\end{tabular}
\caption{Hall A Beam parameters \cite{JLAB-TN-18-022}. }
  \label{Tab:beamparameters}
\end{center}
\end{table}


This paper describes and shows the results of the ARC energy method for the five energy passes, and the elastic energy measurement for the half pass. Then, it reviews the sources of uncertainty for each measurement and its accuracy, and finally, a comparison of the ARC and elastic measurement is done.

\section{ARC Measurement Method}

\begin{figure}[ht]
\includegraphics[width=0.47\textwidth]{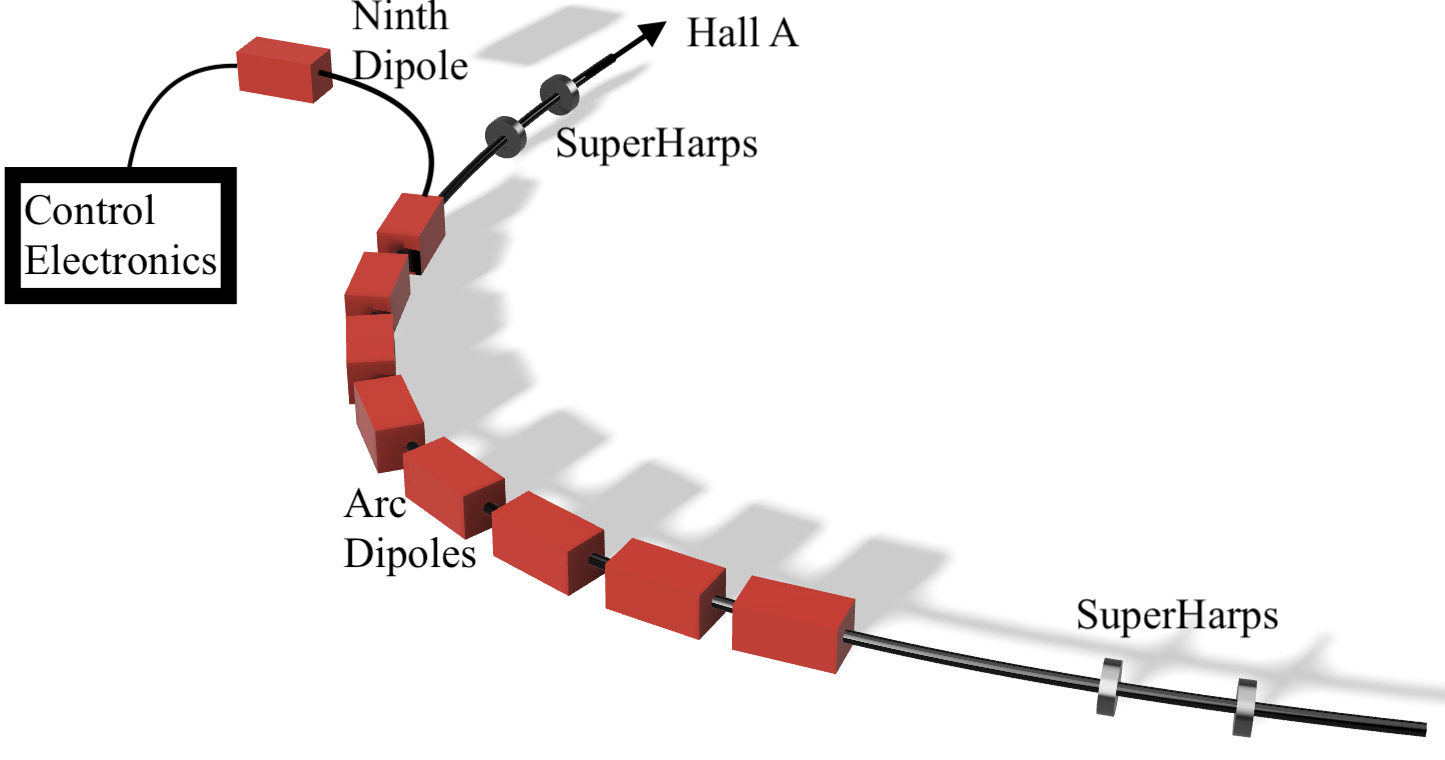}
\caption{ARC beam line section. Eight magnets are used to bend the beam 34.3$^{\circ }$, the ninth magnet is used as reference to perform a direct field measurement.  }
\label{fig:arc}
\end{figure}

The ARC used for the energy measurement comes from the south linac and goes to Hall A, as is layout in Figure \ref{fig:cebaf}. It is a 40 m beam section with  9 dipole magnets, as shown in Figure \ref{fig:arc}. Eight of the magnets are used to deflect the beam, the bending angle is of approximately 34.3$^{\circ }$ and it has been surveyed several times during the past several years, Table \ref{Tab:arcangle} summarizes the results of the surveys. 

The measurement of the bending angle is done by wire scanners called SuperHarps. In practice, the wires of the SuperHarps are moved across the beam path, and when the beam hits the wire, the signal is collected by a photomultiplier tube (PMT).  Based on the signal, the exact position of the beam is reconstructed.  There are two SuperHarps 
in the ARC; one at the entrance and one at the exit of the ARC.  Both are used to 
reconstruct the bend angle.

During the actual operation of the experiments, it is not practical to use the SuperHarps since they do an absolute but intrusive measurement. Instead, beam position monitors (BPMs) are used.  A BPM consists of a cavity with four wire antennas oriented parallel to the electron beam; the radio-frequency (RF) signal from each antenna is converted into a DC signal, which is proportional to the distance between the beam and the antenna.  From these four signals, the position of the beam can be determined~\cite{Barry:1990vh}. The BPMs provide a relative measurement of the position and need regular harp scans from the SuperHarps to be calibrated and properly used.

\begin{table}[ht]
\begin{center}
\begin{tabular}{c c c}
\hline
&  \textbf{Surveyed Arc} & \textbf{Uncertainty}\\
 \textbf{Year} &  \textbf{Angle ($^{\circ} $)} & \textbf{Angle ($^{\circ} $)}	\\ \hline
1999 & 34.302 & 0.001 \\
2001 & 34.295 & 0.001\\
2002 & 34.399 & 0.001 \\
2014 & 34.259 & 0.001 \\
2018 & 34.257 & 0.001 \\        
2021 & 34.259 & 0.001 \\ \hline
\end{tabular}
\caption{Bending angle of the ARC measured by the Jefferson Lab Alignment Group throughout the years.}
\label{Tab:arcangle}
\end{center}
\end{table}

The ARC measurement is based on the fact that the electron moves in a circular trajectory in a magnetic 
field~\cite{Instrumentation}. The momentum of the electrons depends on the magnitude of the magnetic field and its bending angle,

\begin{equation}
E = k \left(\frac{\int \vec{B} \cdot d\vec{l}}{\theta}\right)
\label{eq:arc}
\end{equation}

\noindent where $k = 0.299792$ GeV rad/T-m is the speed of light, $\int \vec{B} \cdot d\vec{l}$ is the field integral of the dipole magnets in T-m, and $\theta$ is the bending angle of the beam in radians. The magnetic field is measured using the ninth dipole, which is located outside the beam line.

The energy measurement can be done either when the beam is in dispersive or achromatic mode. The dispersive mode requires the quadrupoles to be off, and the energy determination will follow Equation \ref{eq:arc}. However, quadrupoles are required for precise alignment and focus of the beam, therefore, corrections to Equation \ref{eq:arc} have to be made. As a consequence, during production runs the quadroples are in used in an achromatic mode, and in some occasions, the quadrupoles are turned off in a dispersive mode for Energy measurements checks. 

\subsection{Determining ARC Field Integral}

\begin{figure}
    \centering
    \includegraphics[width=0.45\textwidth]{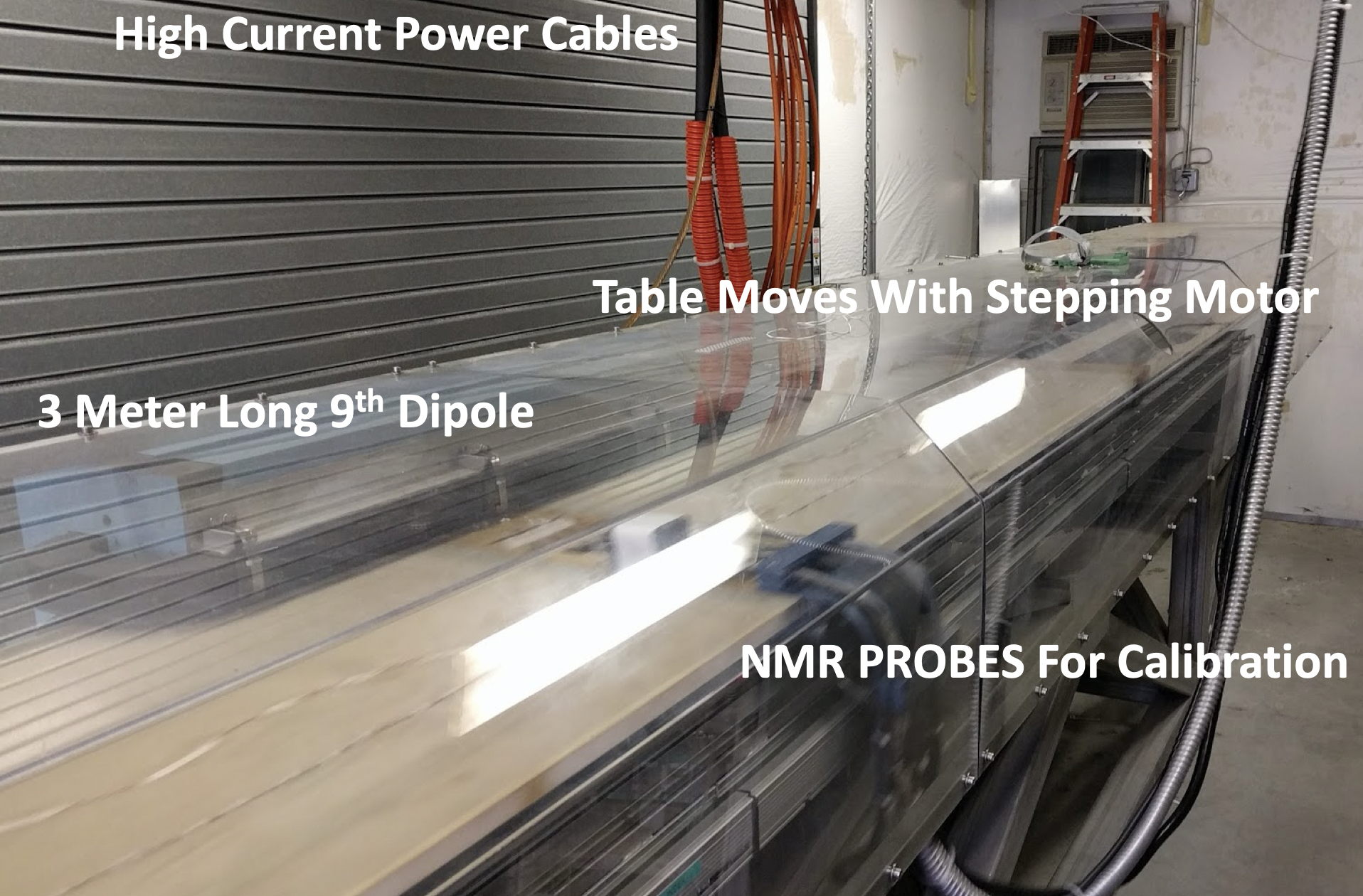}
    \caption{Photo of the ARC field integratal mapping system.  In the background is the 3 meter long 9th dipole which is wired in series with the 8 other dipoles that bring the beam into Hall A.  The mapping table has pickup coils to measuring field gradients and NMR probes for absolute calibrations.}
    \label{fig:mapper}
\end{figure}

The field integral is determined by using an identical 9th dipole magnet located in a building above the beam line as shown in Fig.~\ref{fig:mapper}.  This magnet is powered in series with the other dipoles, and is used as a reference.  It is not possible to measure the field integral of the 8 dipoles in the arc as their gap is fully occupied by the vacuum pipe of the beam.

The fringe effects are accounted in the reference dipole by the use of two translating coils~\cite{Berthot}.  This pair of coils are wired in series and precisely spaced from one another.  As the coils are translated through the 9th dipole's gap, the flux through them is measured.  While these coils give an extremely precise measurement, this measurement is only relative.  In order to achieve an accurate absolute measurement a nuclear magnetic resonance (NMR) probe is used.
	
\begin{figure}[ht]
\begin{center}
\includegraphics[width=0.45\textwidth]{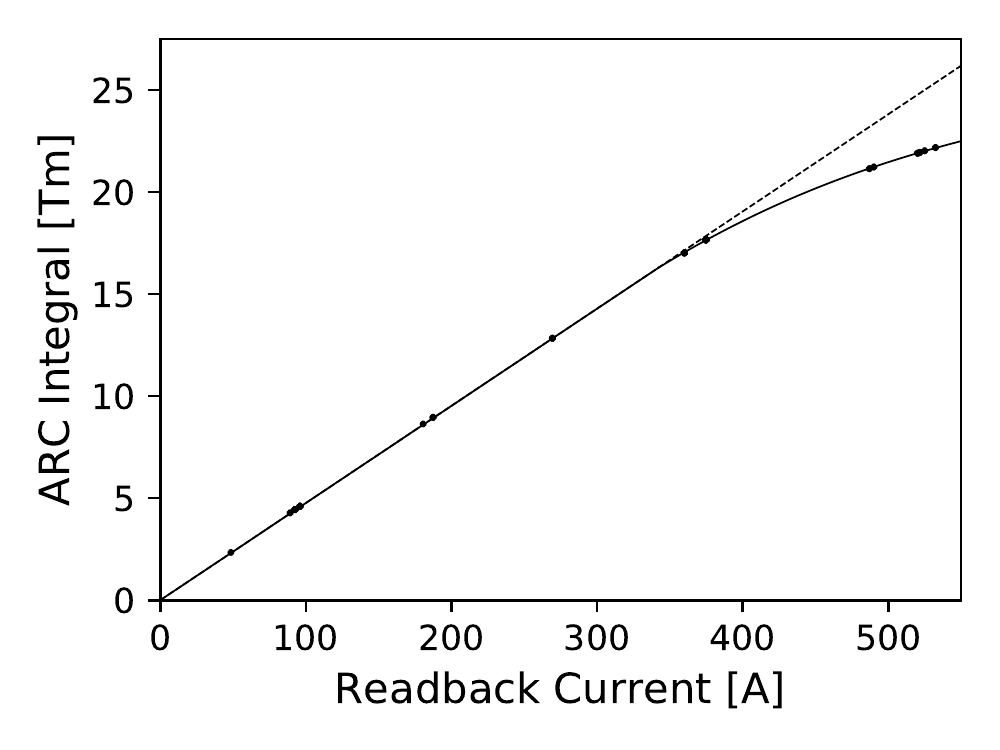}
\caption{ARC integral measurements for various readback currents in the 9th dipole.   At currents below 300~A the system is linear and beyond that begins to saturate.}
\label{fig:arcint}
\end{center}
\end{figure}

Figure \ref{fig:arcint} shows the measured ARC current against the value of the ARC field integral.  The field integral was calculated by mapping the field of the ninth dipole magnet, and measuring the current running through it. It can be seen how the ARC field integral saturates at around 20 T-m where the relationship is no longer linear. These measurements  are applicable in dispersive mode, and serve to provide a value of the ARC field integral based upon the measured ARC current.   

\begin{figure}[ht]
\begin{center}
\includegraphics[width=0.45\textwidth]{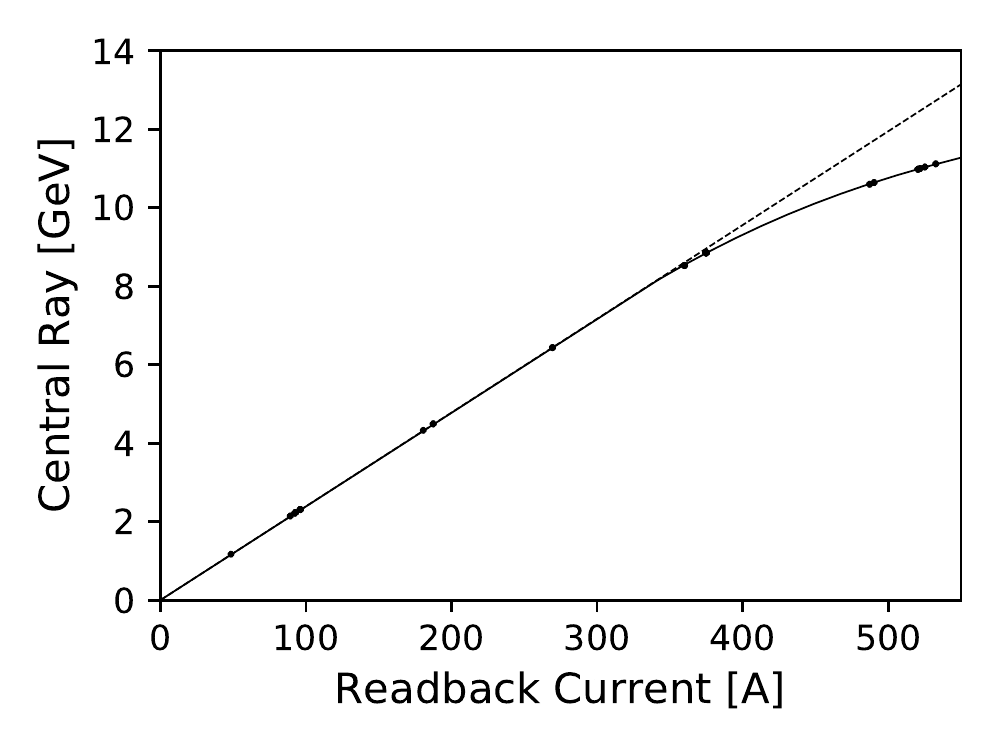}   
\caption{Energy determinations for the ideal central ray through the 8 ARC dipoles.   To minimize the need for orbit corrections, the field in the ARC is abjusted, keeping the beam energy fixed, until by the beam is following the ideal central ray.}
\label{fig:arcint2}
\end{center}
\end{figure}

At low energies a linear fit is an accurate description, but at higher energies 
(and higher ARC current) a rational fit was applied.
	
%
%
%
%

\subsection{HALLA:p and HALLA:dpp}

Another measurement of the beam energy is the HALLA:p variable. In the past the variable was referred to as the ``Tiefenback Energy'', named for its original creator, Michael Tiefenback. The variable currently in use is created via a method derived by Yves Roblin. 
	
The method created by Roblin relies upon the ``Electron Generation and Tracking'' (ELEGANT) program~\cite{ELEGANT}. ELEGANT is a Self-Describing Data Set" (SDDS) compliant program which accesses information about the accelerator from the ``CEBAF Element Database'' (CED). The CED is a database containing many key pieces of information such as magnet field maps~\cite{CED}. This variable is built upon two parts: $p_0$ and HALLA:dpp. $p_0$ is determined using the ARC measurement method and field maps performed during the 6 GeV era. To first order, HALLA:dpp can be approximated by $\partial p$ which is defined below to depend upon the position of the beam measured at the central beam position monitor, 
	
\begin{equation}
    p_{HallA} = p_0 + \partial p
\end{equation}
	
where:
	
\begin{equation}
    \partial p = kx
\end{equation}
	
with k being a constant.  This formula is based upon the fact that lower beam energies will be bent more in the ARC. It is important to note that these formulae are merely the first order terms of the orbit correction performed by Roblin's calculation. There are higher order terms which make further orbit corrections. Figure 3 demonstrates that these higher order corrections amount to less than 2 MeV of the beam energy measurement. Thus, an investigation into these higher order terms is not warranted for this analysis.
	
The primary assumption made in the creation of the HALLA:p variable is that the value for $p_0$ is correct. It is important to note that HALLA:dpp depends very weakly on this value of p0. Table 4 demonstrates a discrepancy between this HALLA:p variable and ARC measurements taken by Doug Higinbotham.  Since measurements made via Higinbotham's method have shown agreement with other measurement methods in the 6 GeV era, this discrepancy must point to an issue with the HALLA:p calculation of $p_0$.

\section{Elastic Energy Measurement}

During the E12-11-112 experiment, the accelerator was able to send half pass to Hall A by using only the North Linac, since the South Linac was explicitly turned off. The Left High Resolution Spectrometer (LHRS) was positioned with an opening angle of 17.009$^{\circ }$ and with a central momentum of 1.1282 GeV. 

The basic components of the LHRS~\cite{Alcorn:2004sb} are three superconducting quadrupoles (Q) and one superconducting dipole (D) in a QQDQ configuration. The quadrupoles align the scattered electrons while the dipole determines their momentum. The central momentum of the LHRS was calculated using NMR measurements of the magnets as given by Liyanage~\cite{JLab-TN-01-049}. The momentum was chosen such that both Elastic and Quasi-Elastic peaks were accessible for the $^{3}$H and $^{3}$He gas targets. Furthermore,  $^{1}$H was also available, and using these three gas targets the beam energy was measured. 

The sealed gas target cells were used in a modular low pressure system as described 
in~\cite{SANTIESTEBAN2019351}. The length of the target was 25 cm and for the
purposes of analysis only data from  the center of the target $\pm$8\,cm was 
selected, in order to avoid contamination coming from the aluminum end caps of 
$<$4 mm thick as shown in~\cite{SANTIESTEBAN2019351}.  The nominal current provided
for this data  was of 5\,$\mu$A, and only events with $\pm$1.5\,$\mu$A were selected 
order to avoid any drift or trip in the beam. 

After passing the magnets, the scattered electrons go through two Vertical Drift 
Chambers (VDCs), where the electrons ionize the gas inside the chambers, and with 
the information collected, the position and angle of the trajectory of the electrons are found. Then, the trigger
scintillators s0 and s2m are used to record the selected candidates. Due to the high 
rate of the kinematics, the prescaler chosen for the trigger was 25. The Cherenkov 
detector filled with CO$_{2}$ between the trigger scintillator planes identify the 
electrons with 99\% efficiency and has a threshold for pions of 4.8\,GeV. Finally, 
the preshower and shower lead glasses blocks induce a cascade of pair production and 
bremsstrahlung radiation from energetic particles, which are used for the measurement
of the energy of the electrons.  In order to select a clean sample of electrons, we 
require a single track in the VDC chamber, E/p $>$ 0.7, Cherenkov signal above the
pion contamination ($>$1500), an out-of-plane angle $\pm$35\,mrad and an in-plane angle $\pm$30\,mrad. 

In the elastic scattering, the energy of the beam and the energy of the scattered electrons is related by: 

\begin{equation}
    E= \frac{E_{i}-E_{loss1}}{1+\frac{(E_{i}-E_{loss1})sin^{2}(\theta /2)}{M_{t}}} + E_{loss2}
    \label{eq:elastic}
\end{equation}

where E is the beam energy, E$_{i}$ is the energy of the scattered electrons, $\theta$ is the scattering angle, E$_{loss1}$ and E$_{loss2}$ are the energy losses before and after the scattering, and $M_{t}$ is the mass of the target. 

\begin{figure*}[htb]
\centering
	\includegraphics[width=\textwidth]{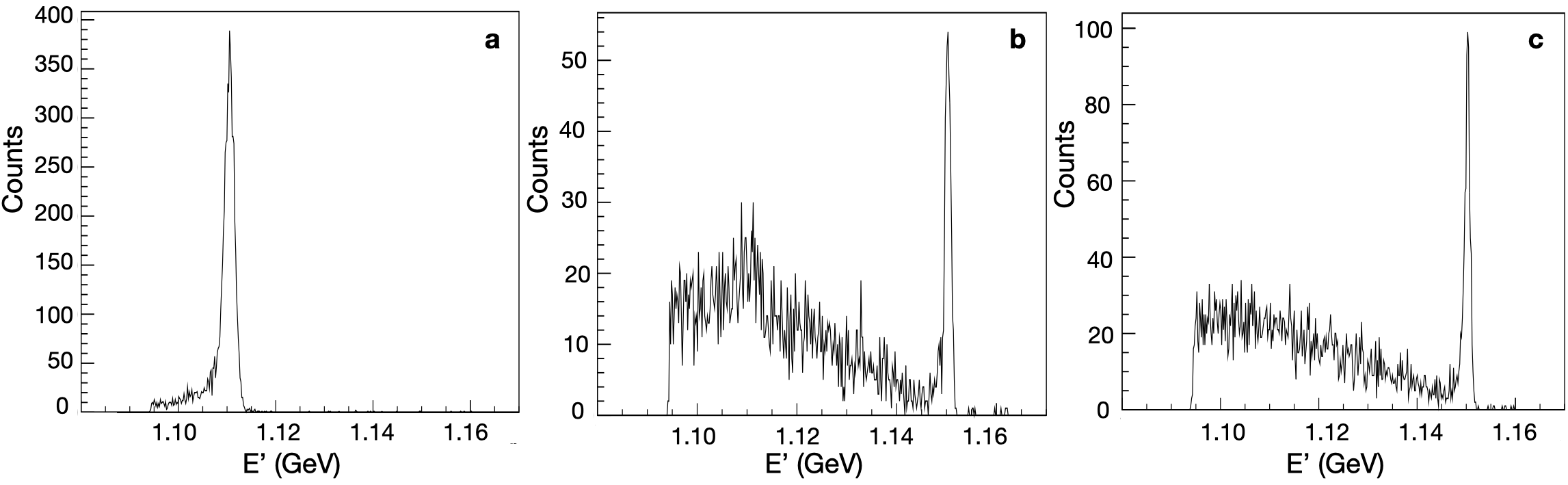}
\caption{Scattered energy of the electrons withing an angle of 17$^{\circ}$ $\pm$ 0.1$^{\circ}$, for the elastic peak in $^{1}$H (a), and the Quasi-Elastic peaks for $^{3}$H (b) and $^{3}$He (c).  }
\label{fig:momentum}
\end{figure*}

Using the Equation~\ref{eq:elastic} the beam energy can be calculated, if the scattering angle is the same and the scattered energy is known. Therefore, the measurement was done by selecting particles with a scattering angle of 17$^{\circ}$ $\pm$ 0.1$^{\circ}$. Figure~\ref{fig:momentum} shows the distribution for the scattered energy after the energy loss correction.

To measure the scattered energy of the electrons, a Gaussian fit was done and the
results are shown in the Table~\ref{tab:mom}. The  scattered energy ($E'$) and its 
uncertainty ($dE'$) was measured for each target. 

\begin{table}[ht]
    \centering
    \begin{tabular}{c|c|c}
         Target & E'(GeV) & dE'(GeV)   \\
         \hline 
         $^{1}$H & 1.1104 & 6.5x10$^{-04}$\\
         $^{3}$H & 1.15029 & 7.1x10$^{-04}$\\
         $^{3}$He & 1.15022 & 8.2x10$^{-04}$ \\ 
    \end{tabular}
    \caption{Scattered electron energies for$^{1}$H , $^{3}$H and $^{3}$He. }
    \label{tab:mom}
\end{table}

Finally, in order to account for the systematic uncertainties, several factors were 
studied. The summary is presented in Table \ref{tab:systematics}, the LHRS central
momentum represents the uncertainty given by the NMR calculation in ~\cite{JLab-TN-01-049} 
and the angle resolution represents the cut dependence of the angle chosen to measure the beam energy.

\begin{table}[ht]
    \centering
    \begin{tabular}{c|c}
        LHRS central momentum  & 4x10$^{-4}$   \\
        \hline
        Angle Resolution & 8x10$^{-5}$
    \end{tabular}
    \caption{Systematic Uncertainties}
    \label{tab:systematics}
\end{table}{}

\begin{figure*}[ht]
\centering
	\includegraphics[width=\textwidth]{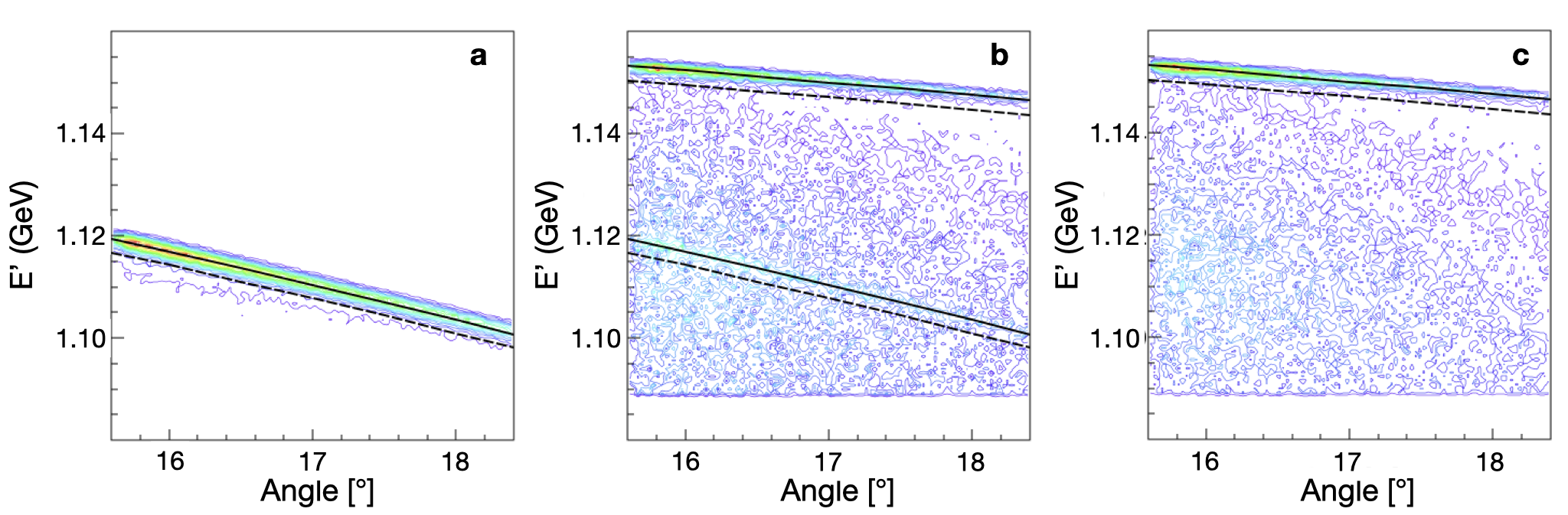}
\caption{E' with respect to the  Scattering angle ($\theta )$ for $^{1}$H (a), $^{3}$H (b) and $^{3}$He (c). The dashed lines corresponds to the calculations from the energy reported by the accelerator and solid lines are from the energy measured in this work} 
\label{fig:2D}
\end{figure*}

The beam energy measured using this technique is 1171.3 $\pm$ 0.6~MeV. While the energy reported by the accelerator was of 1168 $\pm$ 1~MeV. As a sanity check, Figure \ref{fig:2D} shows the scattered electrons energy with respect to the scattering angle, the black and red lines correspond to the calculated E' using the accelerator and the measured energy respectively. It shows that the measured energy reproduces the elastic data,

\section{Spin Precession}

Another technique for cross checking the beam energy is to use the beam energy to make calculations of the electron spin precession from the injector to the experimental hall~\cite{Higinbotham:2009ze,Higinbotham:2014yua}.   These calculations are used to predict the Wein angle setting at the injector to provide full polarization to a hall.   This calculation can easily be check by making polarizaton measurements with a Moller polarimeter in the hall in a technique known as a spin dance~\cite{Grames:2004mk}.

\begin{figure}[htb]
    \centering
    \includegraphics[width=\linewidth]{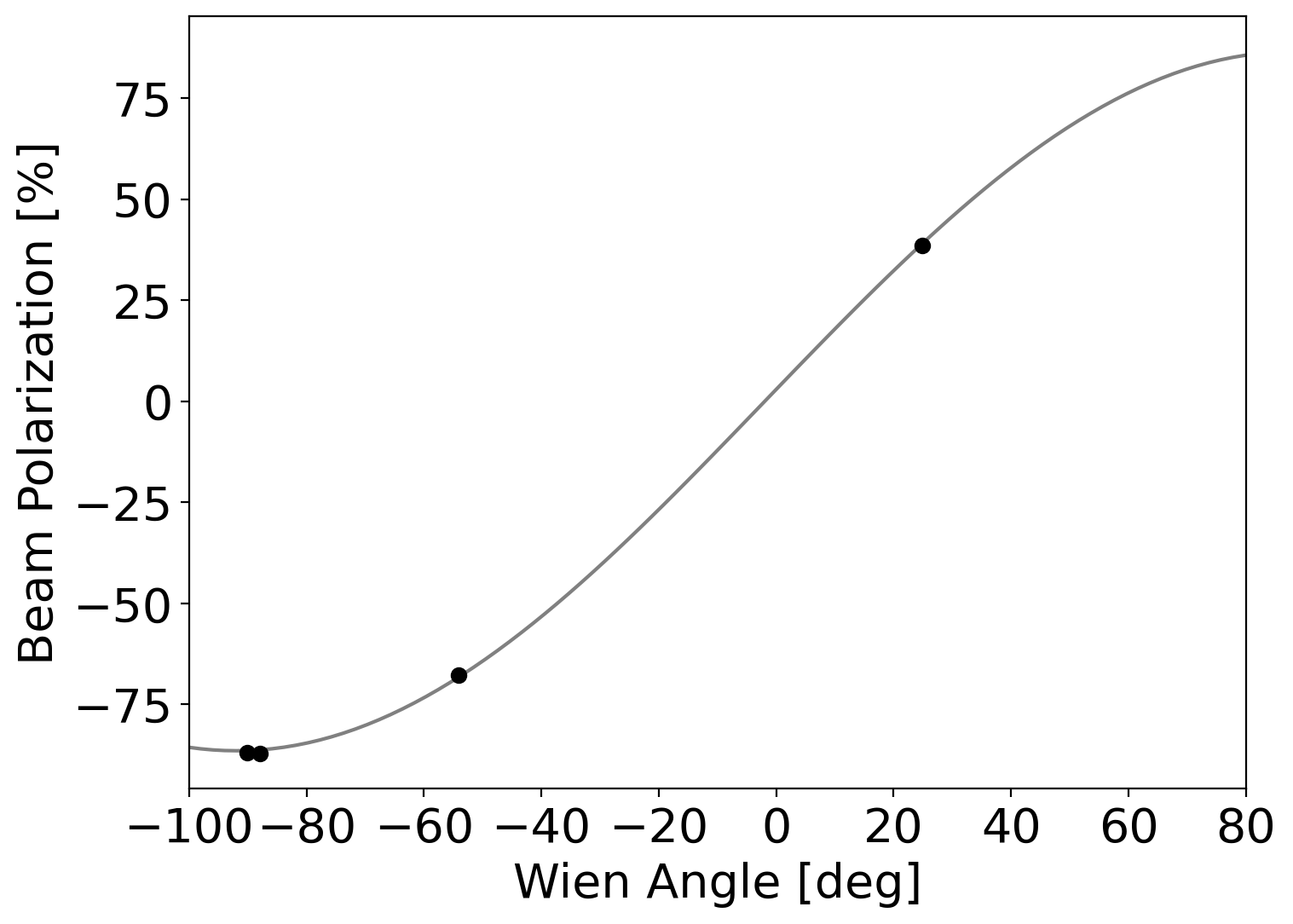}
    \caption{Moller polarimeter results for spin dance in November of 2016. The points are the experimental data and the line a cosine fit. From the fit, the Wien Angle corresponding to the maximum longitudinal polarization is 86.48 $\pm$ 0.1204 degrees which is in agreement with the peak angle of 85 $\pm$ 2 degrees that was predicted using the ARC energy measurement to calculate the expected maximum. }
    \label{fig:my_dance}
\end{figure}

Using the information from the ARC energy measurements, it was predicted the Wein should be set for 85 degrees.   Other methods, based on estimates of the beam energy using field maps of the accelerator arc magnets predicted smaller angles.   This results, as shown in Fig.~\ref{fig:my_dance} as well as subsequent spin dance, confirmed that the ARC system is providing an accurate absolute energy and that there was a small offset in the accelerator's real time monitoring calculation.

\section{Real Time Beam Energy}

As has been shown in this manuscript, the ARC energy measurement system can provide an excellent instantaneous measurements of the electron beam energy; but for experiments, one needs to know the beam energy over the course many days and sometimes even many months.

For this, we make use of use a relative energy that is determined using the setting of the Hall A ARC Bdl and beam position monitors located at the beginning, middle and end of the ARC.   While this system provides a real time, relative measurement, it requires the results from another system for the absolute calibration.

During the 6 GeV era, this calibration was done using the ARC energy system as well as an eP elastic scattering system~\cite{Alcorn:2004sb}.   This was done again during the 12 GeV era by simply cross calibrating the real time system against the ARC energy results.   What we have found is that the real time monitor generally needs to be multiplied by 1.003 to give the true energy except at the highest energy where serendipitously the synchrotron radiation, which grows at $E^4$, reduces the energy bring the correction factor back to near unity.

\section {Summary }

In summary, the Hall A ARC mapper system has been used to determine the absolute beam energy of the CEBAF accelerator after the 12 GeV upgrade and to re-calibrate the real time energy monitoring system. To ensure the accuracy of these results they were cross checked against elastic scattering in the Hall A high resolution spectrometers as well as with a spin dance where the Hall A ARC energy result was used to correctly predict the Wein angle setting for providing maximal longitudinal polarization. 

\section{Acknowledgments}

We wish to thank the accelerator staff of the Thomas Jefferson National Accelerator Facility without them these measurements would not have been possible.   We would also like to thank Michael Tiefenback and Yves Roblin for many very useful discussions.
This material is based upon work supported by the U.S. Department of Energy, Office of Science, Office of Nuclear Physics under contract DE-AC05-06OR23177.

\section*{References}

\bibliographystyle{elsarticle-num} 
\bibliography{references.bib}

\begin{thebibliography}{10}
\expandafter\ifx\csname url\endcsname\relax
  \def\url#1{\texttt{#1}}\fi
\expandafter\ifx\csname urlprefix\endcsname\relax\def\urlprefix{URL }\fi
\expandafter\ifx\csname href\endcsname\relax
  \def\href#1#2{#2} \def\path#1{#1}\fi

\bibitem{Leemann:2001dg}
C.~Leemann, D.~Douglas, G.~Krafft, {The Continuous Electron Beam Accelerator
  Facility: CEBAF at the Jefferson Laboratory}, Ann. Rev. Nucl. Part. Sci. 51
  (2001) 413--450.
\newblock \href {http://dx.doi.org/10.1146/annurev.nucl.51.101701.132327}
  {\path{doi:10.1146/annurev.nucl.51.101701.132327}}.

\bibitem{Grames:2010zz}
J.~Grames, D.~W. Higinbotham, H.~E. Montgomery, {Thomas Jefferson National
  Accelerator Facility}, Nucl. Phys. News 20N3 (2010) 6--13.
\newblock \href {http://dx.doi.org/10.1080/10619127.2010.506115}
  {\path{doi:10.1080/10619127.2010.506115}}.

\bibitem{828507}
F.~{Kircher}, J.~{Fabre}, F.~{Gougnaud}, M.~{Humeau}, R.~{Leboeuf},
  Y.~{Lussignol}, J.~{Marroncle}, G.~{Matichard}, D.~{Marchand}, J.~C.
  {Sellier}, P.~{Vernin}, C.~{Veyssiere}, High accuracy field integral
  measurement for tjnaf beam energy determination, IEEE Transactions on Applied
  Superconductivity 10~(1) (2000) 1427--1430.

\bibitem{marchand:tel-00298382}
D.~Marchand, \href{https://tel.archives-ouvertes.fr/tel-00298382}{{Calcul des
  corrections radiatives {\`a} la diffusion compton virtuelle. Mesure absolue
  de l'{\'e}nergie du faisceau d'{\'e}lectrons de Jefferson Lab. (Hall A) par
  une m{\'e}thode magn{\'e}tique : projet ARC}}, Theses, {Universit{\'e} Blaise
  Pascal - Clermont-Ferrand II} (Apr. 1998).
\newline\urlprefix\url{https://tel.archives-ouvertes.fr/tel-00298382}

\bibitem{Berthot:1999jp}
J.~Berthot, P.~Vernin, {Beam energy measurement in Hall-A of CEBAF}, Nucl.
  Phys. News 9N4 (1999) 12--16.
\newblock \href {http://dx.doi.org/10.1080/10506899909411146}
  {\path{doi:10.1080/10506899909411146}}.

\bibitem{Grames:2004mk}
J.~Grames, et~al., {Unique electron polarimeter analyzing power comparison and
  precision spin-based energy measurement}, Phys. Rev. ST Accel. Beams 7 (2004)
  042802, [Erratum: Phys.Rev.ST Accel.Beams 13, 069901 (2010)].
\newblock \href {http://dx.doi.org/10.1103/PhysRevSTAB.7.042802}
  {\path{doi:10.1103/PhysRevSTAB.7.042802}}.

\bibitem{Higinbotham:2009ze}
D.~Higinbotham, {Electron Spin Precession at CEBAF}, AIP Conf. Proc. 1149~(1)
  (2009) 751--754.
\newblock \href {http://arxiv.org/abs/0901.4484} {\path{arXiv:0901.4484}},
  \href {http://dx.doi.org/10.1063/1.3215753} {\path{doi:10.1063/1.3215753}}.

\bibitem{Higinbotham:2013pgc}
D.~W. Higinbotham, {Using Polarimetry To Determine The CEBAF Beam Energy}, PoS
  PSTP2013 (2013) 014.
\newblock \href {http://dx.doi.org/10.22323/1.182.0014}
  {\path{doi:10.22323/1.182.0014}}.

\bibitem{Berthot}
J.~Berthot, P.~Vernin, {Beam energy measurement in Hall-A of CEBAF}, Nucl.
  Phys. News 9N4 (1999) 12--16, \url{http://www.jlab.org}.
\newblock \href {http://dx.doi.org/10.1080/10506899909411146}
  {\path{doi:10.1080/10506899909411146}}.

\bibitem{JLAB-TN-18-022}
J.~Benesch, A.~Bogacz, A.~Freyberger, Y.~Roblin, T.~Satogata, R.~Suleiman,
  M.~Tiefenback, 12 gev cebaf beam parameter tables, Tech. Rep. JLAB-TN-18-022,
  Thomas Jefferson National Accelerator Facility (2018).

\bibitem{Barry:1990vh}
W.~Barry, J.~Heefner, J.~Perry, {Electronics systems for beam position monitors
  at CEBAF}, AIP Conf. Proc. 229 (1991) 48--74.
\newblock \href {http://dx.doi.org/10.1063/1.40758}
  {\path{doi:10.1063/1.40758}}.

\bibitem{Instrumentation}
J.~Alcorn, et~al., Basic instrumentation for {Hall A at Jefferson Lab}, Nuclear
  Instruments and Methods in Physics Research Section A: Accelerators,
  Spectrometers, Detectors and Associated Equipment 522~(3) (2004) 294 -- 346.
\newblock \href {http://dx.doi.org/https://doi.org/10.1016/j.nima.2003.11.415}
  {\path{doi:https://doi.org/10.1016/j.nima.2003.11.415}}.

\bibitem{ELEGANT}
M.~Borland, T.~Berenc, Users manual for elegant (2019).

\bibitem{CED}
T.~Larrieu, M.~Joyce, C.~Slominski, Design and implementation of the cebaf
  element database, Proceedings of ICALEPCS2011.

\bibitem{Alcorn:2004sb}
J.~Alcorn, et~al., {Basic Instrumentation for Hall A at Jefferson Lab}, Nucl.
  Instrum. Meth. A 522 (2004) 294--346.
\newblock \href {http://dx.doi.org/10.1016/j.nima.2003.11.415}
  {\path{doi:10.1016/j.nima.2003.11.415}}.

\bibitem{JLab-TN-01-049}
N.~Liyanage, Spectrometer constant determination for the hall-a resolution
  spectrometer pair, Tech. Rep. JLAB-TN-18-022, Thomas Jefferson National
  Accelerator Facility (2001).

\bibitem{SANTIESTEBAN2019351}
S.~Santiesteban, et~al., {Density Changes in Low Pressure Gas Targets for
  Electron Scattering Experiments}, Nucl. Instrum. Meth. A 940 (2019) 351--358.
\newblock \href {http://arxiv.org/abs/1811.12167} {\path{arXiv:1811.12167}},
  \href {http://dx.doi.org/10.1016/j.nima.2019.06.025}
  {\path{doi:10.1016/j.nima.2019.06.025}}.

\bibitem{Higinbotham:2014yua}
D.~W. Higinbotham, {Using Polarimetry To Determine The CEBAF Beam Energy}, PoS
  PSTP2013 (2013) 014.
\newblock \href {http://dx.doi.org/10.22323/1.182.0014}
  {\path{doi:10.22323/1.182.0014}}.

\end{thebibliography}

\end{document}